\documentstyle[12pt]{article}
\newcommand {\cA}{{\cal A}}
\newcommand {\cB}{{\cal B}}

\newcommand {\cD}{{\cal D}}

\newcommand {\cK}{{\cal K}}

\newcommand {\cP}{{\cal P}}
\newcommand {\cQ}{{\cal Q}}
\newcommand {\cR}{{\cal R}}

\newcommand {\cV}{{\cal V}}

\def\a{\alpha}
\def\b{\beta}

\def\d{\delta}

\def\f{\phi}
\def\g{\gamma}
\def\G{\Gamma}

\def\j{\psi}

\def\l{\lambda}
\def\m{\mu}

\def\q{\theta}

\def\U{\Upsilon}

\newcommand{\ad}{{\dot{\alpha}}}                           %new
\newcommand{\bd}{{\dot{\beta}}}                            %new
\newcommand{\ve}{\varepsilon}                            %new
\newcommand{\cDB}{{\bar\cD}}                            %new

\renewcommand{\aa}{{\a\ad}}

\newcommand{\kl}{{\a(k)\ad(l)}}

\renewcommand{\ss}{{\a(s)\ad(s)}}
\newcommand{\rr}{{\a(s-1)\ad(s-1)}}

\newcommand{\rs}{{\a(s-1)\ad(s)}}

\newcommand{\ts}{{\a(s+1)\ad(s)}}

\newcommand{\hf}{\frac12}

\newcommand{\sect}[1]{\setcounter{equation}{0}\section{#1}}

\newcommand{\be}{\begin{equation}}
\newcommand{\ee}{\end{equation}}
\newcommand{\bea}{\begin{eqnarray}}
\newcommand{\eea}{\end{eqnarray}}
\newcommand{\non}{\nonumber}

\begin{document}
\begin{titlepage}

\begin{flushright}
UMDEPP 97-28 \\
TSU--QFT--15/96 \\
hep-th/9611193 \\
\end{flushright}

\begin{center}
\large{{\bf Towards a Unified Theory of Massless Superfields \\
of All Superspins} } \\
\vspace{1.0cm}

\large{S. James Gates Jr.\footnote{ E-mail: gates@umdhep.umd.edu.}  } \\

\footnotesize{{\it Department of Physics, University of Maryland \\
College Park, MD 20742-4111, USA}} \\
\vspace{0.5cm}

\large{Sergei M. Kuzenko\footnote{ E-mail: kuzenko@phys.tsu.tomsk.su.}
and Alexander G. Sibiryakov\footnote{ E-mail: sib@phys.tsu.tomsk.su.}
 } \\

\footnotesize{{\it Department of Quantum Field Theory, Tomsk State
University \\
Lenin Ave. 36, Tomsk 634050, Russia} \\
 }
\end{center}
\vspace{1.5cm}

\begin{abstract}
We describe the {\it {universal}} linearized action for massless superfields
of all superspins in $N=1$, $D=4$ anti-de Sitter superspace as a gauge theory
of unconstrained superfields taking their values in the commutative algebra
of analytic functions over a one-sheeted hyperboloid in
${\bf R}^{3,1}$.  The action is invariant under $N=2$ supersymmetry
transformations which form a closed algebra off the mass-shell.
\end{abstract}
\vspace{15mm}

\begin{flushleft}
November 1996
\end{flushleft}

\vfill
\null
\end{titlepage}
\newpage
\setcounter{footnote}{0}

\sect{Introduction}

${~~~~}$Off-shell superfield realizations of $N=1$, $D=4$ higher superspin
massless multiplets were given previously \cite{ks1} (see also \cite{bk})
for Poincar\'e supersymmetry and then extended to the case of anti-de Sitter
(AdS) supersymmetry \cite{ks2}. The models obtained (both with integer as
well as half-integer superspins) naturally form two dually equivalent series.
Any cursory glance at the structure of these models suggests that there
should exist a universal formulation in which massless models of all
superspins occur as special cases.

The use of such a formulation may be indispensable for constructing a
consistent theory of interacting massless (super)fields of all
(super)spins. In this respect it is worth mentioning a totally
consistent system of equations for interacting massless fields
of all spins, including the gravitational field, constructed by
Vasiliev \cite{vas1} and based on a geometric generating description
for (properly generalized) higher spin gauge massless models in
the AdS space \cite{fr,vas}. Unfortunately, up to now it remains
unknown whether these equations can be derived from an action
functional.

In the present paper we propose a generating formulation (wherein each
superspin enters with multiplicity one) for the combined action of free
massless superfields of all superspins in $N=1$, $D=4$ AdS superspace
$z^M = (x^m,\q^{\m}, \bar{\q}_{\dot{\m}})$.
We realize this model as a gauge theory of unconstrained
superfields $\varphi (z^M, q^a)$ over the AdS superspace and
analytic in a four-vector variable $q^a$ constrained by
$q^aq_a=1$.  Usual higher tensor representation superfields emerge as
coefficients in power series expansions in $q$. 

{}From a geometrical point of view, our formulation is based
on enlarging the AdS superspace by 
a one-sheeted hyperboloid in ${\bf R}^{3,1}$ which is parametrized
by $q^a$.
The bosonic sector of the final superspace can be realized as a
homogeneous space of $O(3,2)$, i.e. the symmetry group of
the AdS space. An $O(3,2)$ covariant realization involves
five-vectors $y^\cA$ and $q^\cA$, $\cA = 5,0,1,2,3$, under
the constraints
\be
y^\cA y_\cA = - r^2 ~~, \qquad
y^\cA q_\cA = 0 ~~, \qquad q^\cA q_\cA = 1 ~~~,
\label{cov}
\ee
where the indices are contracted with the use of
the metric $\eta_{\cA\cB} =diag(--+++)$.
The constraint variables $y^\cA = y^\cA(x^m)$ parametrize the AdS space,
with $-12r^{-2}$ the corresponding curvature.
A remarkable feature of the manifold introduced is that spacetime
(parametrized by $y$) and internal space (parametrized by $q$)
originate, in a sense, on equal footing.
This becomes obvious by rewriting (\ref{cov}) as follows
\be
q_\pm q_{\pm}= 0 ~~, \qquad q_{+}q_{-} = 1 ~~,
\label{cov1}
\ee
where $q^\cA_\pm = (r^{-1}y^\cA \pm q^\cA)/\sqrt2$.
The constraints remain unchanged under $SO(1,1)$ and discrete
transformations
\bea
 q_+ \rightarrow \l q_+ ~~,  & \qquad & q_- \rightarrow
\l^{-1} q_- ~~, \qquad \l \neq 0 ~~, \\
 q_+ \rightarrow  q_- ~~, & \qquad & q_- \rightarrow  q_+ ~~,
\eea
which commute with the AdS transformations.

  Our present formulation for the combined action of massless multiplets
of {\it all} superspins is described by two real scalar superfields
$X(z,q)$, $Y(z,q)$ and a complex one $U(z,q)$, and all gauge
transformations arise from a complex parameter $\ve (z,q)$.  The most
surprising thing is that the model obtained admits a considerable
simplification by imposing the algebraic gauge condition $Y=0$ which
leads to an elegant formulation. What is more,
the second real superfield $X$ is auxiliary and can be
eliminated without explicit supersymmetry breaking. Another remarkable
feature of the proposed model is that it possesses an irreducible gauge
algebra instead of the infinitely reducible gauge transformations
present in the original higher superspin models \cite{ks2}.
The approach under consideration enables us to combine all the Killing
tensor superfields of the AdS superspace within a single superfield
$\U(z,q)$ satisfying a simple equation. This observation presents an
exceptional possibility for the covariant study of the
infinite-dimensional global symmetry of the higher-superspin actions
which is likely to be described  by the superalgebra proposed in
\cite{fv}.

\sect{Arbitrary Superspin Massless Models in the AdS Superspace}

${~~~~}$ The $N=1$, $D=4$ AdS superspace \cite{ads}, with coordinates
$z^M = (x^m,\q^{\m}, \bar{\q}_{\dot{\m}})$, 
is specified by the algebra of covariant derivatives
$\cD_A = (\cD_a, \cD_\a, \cDB^\ad)$\footnote{Our two-component
notations and conventions coincide with those adopted in \cite{bk,wb}.
We consider only Lorentz tensors
symmetric in their undotted indices and separately in dotted ones.
A tensor of type $(k,l)$ with $k$ undotted and $l$ dotted indices
can be equivalently represented as
$\psi(k,l) \equiv \psi_\kl   \equiv \psi_{\a_1 \ldots
\a_k\ad_1\ldots \ad_l} = \psi_{(\a_1 \cdots \a_k)(\ad_1\cdots
\ad_l)}$.
Following Ref. \cite{vas}, we assume that the indices,
which are denoted by one and the same letter,
should be symmetrized separately with respect to upper and
lower indices; after the symmetrization, the maximal possible number
of the upper and lower indices denoted by the same letter
are to be contracted. In particular
$\phi_{\a(k)} \psi_{\a(l)}
\equiv \phi_{(\a_1\cdots\a_k} \psi_{\a_{k+1}\cdots\a_{k+l})}$
and $\xi^\a \phi_{\a(k)} \equiv  \xi^\b \phi_{(\b\a_1\cdots\a_{k-1})}$.
Given two tensors of the same type, their contraction is denoted by
$\phi \cdot \psi \equiv \phi^\kl \psi_\kl$.}
\bea
\{{\cD}_\a,\cDB_{\ad}\}&=&-2i{\cD}_{\a\ad}~~~, \qquad \;
[{\cD}_{\a\ad},{\cD}_{\b\bd}]=-2\bar\mu \mu
(\varepsilon_{\a\b} \bar M_{\ad\bd} +
\varepsilon_{\ad\bd} M_{\a\b}) ~~~, \non \\
\{ {\cD}_\a,{\cD}_\b\}&=&-4\bar\mu M_{\a\b} ~~~, \qquad
[{\cD}_\a,{\cD}_{\b\bd}]=i\bar\mu
\varepsilon_{\a\b} \cDB_{\bd} ~~~, \non \\
\{ {\cDB}_\ad,{\cDB}_\bd\}&=&4\m \bar{M}_{\ad\bd} ~~~, \qquad \;\;\;
[{\cDB}_\ad,{\cD}_{\b\bd}]=-i\m
\varepsilon_{\ad\bd} \cD_{\b} ~~~.
\label{1}
\eea
Here $M$, $\bar M$ denote the Lorentz generators, the non-zero constant $\m$
of unit mass dimension determines the torsion (and hence the curvature)
of the AdS superspace.  The algebra of covariant derivatives is invariant
under superspace general coordinate transformations and local Lorentz
rotations
\be
\d\cD_A = [\cK, \cD_A] ~~, \qquad
\cK = -\hf k^\aa \cD_\aa + (k^\a \cD_\a
+ k^{\a(2)} M_{\a(2)} + {\rm c.c.})
= \bar\cK ~~,
\label{2}
\ee
with superfield parameters $k^\aa$, $k^\a$, $k^{\a(2)}$.
The covariant derivatives remain unchanged, $[\cK, \cD_A]=0$,
provided the parameters $k^A$, $k^{\a(2)}$ are constrained by
\bea
&& k_\a = \frac i8 \cDB^\ad k_\aa ~~, \quad k_{\a(2)} = \cD_\a k_\a
~~,
\label{3}\\
&&\cDB_\ad k_\aa =  \cD^\a \cDB^\ad k_\aa = 0 ~~.
\label{4}
\eea
Eqs. (\ref{2})--(\ref{4}) define a Killing supervector $\cK$ of the AdS
superspace. The set of all Killing supervectors is known to form the
superalgebra $osp(1,4)$.

Now, we recall some results of \cite{ks2} on the Lagrangian
realization of higher-superspin massless multiplets in the AdS
superspace. An important role in this approach is played by
``transversal'' and ``longitudinal'' linear superfields. A complex
superfield $\G(k,l)$ subject to the constraint
\bea
\cDB^{\ad} \Gamma_{\a(k)\ad(l)} = 0 ~~, \;\;\; \Leftrightarrow\;\;\;
(\bar{\cP}  - l - 2) \G(k,l) &=& 0 ~~, \qquad l > 0~~; \non \\
(\bar{\cP}  -  2) \G(k,0) &=& 0 ~~, \qquad l = 0 ~~;
\label{5}
\eea
is defined as a transversal linear superfield; a complex superfield
$G(k,l)$ subject to the constraint
\be
\cDB_{\ad} G_{\a(k)\ad(l)} = 0~~, \;\;\; \Leftrightarrow\;\;\;
(\bar{\cP} + l) G(k,l) = 0 ~~.
\label{6}
\ee
is defined as a longitudinal linear superfield. Here we have used the
notations
\be
\cP = \frac{1}{2\bar{\m}}\cD^2 = \frac{1}{2\bar{\m}}\cD^\a \cD_\a ~~, \qquad
\bar{\cP} = \frac{1}{2\m}\cDB^2 = \frac{1}{2\m}
\cDB_\ad \cDB^\ad ~~.
\label{7}
\ee
For $l=0$ (\ref{6}) coincides with the chirality constraint.

For each superspin value greater 3/2, there are known \cite{ks2} exactly
two dually equivalent superfield formulations called transversal and
longitudinal. In the present paper we make use of the transversal
formulation for a half-integer superspin $s+1/2$, $s \geq 1$, and of
the longitudinal formulation for an integer superspin $s$, $s \geq 1$.
The dynamical superfield variables in these cases take the forms
\bea
\cV^\bot_{s+\hf} &=& \{\, H(s,s), \G(s-1,s-1), \bar\G(s-1,s-1)\, \}
~~,\qquad s \geq 1 \label{10} \\
\cV^\|_s &=& \{\, H'(s-1,s-1), G(s,s), \bar G(s,s)\, \} ~~, \qquad
\qquad \;\;\; s\geq 1
\label{11}
\eea
where $H$ and $H'$ are real unconstrained, $\G$ transversal linear and $G$
longitudinal linear superfields. The transversal superspin-$(s+1/2)$ action
reads
\bea
S^\bot_{s+\hf}&=&(-1)^s \hf \displaystyle\int {\rm d}^8z E^{-1}\Big\{
~{1\over8} H^\ss{\cD}^\b (\cDB^2-4\mu){\cD}_\b H_\ss {~~~~~~~~~~~~~~~~
~~~~~~~}\cr
& &{~~~~~~~~~~~~~~}+H^\ss \big({\cD}_\a \cDB_{\ad} \G_\rr -
\cDB_{\ad}{\cD}_\a \bar\G_\rr\big)\cr
& &{~~~~~~~~~~~~~~}+\displaystyle\frac{s^2}{2}\bar\mu\mu H\cdot H +
2\bar\G\cdot\G +\frac{s+1}{s}(\G\cdot\G +
\bar\G\cdot\bar\G)~\Big\} ~~,
\label{12}
\eea
and the longitudinal superspin-$s$ action is given by
\bea
S^\|_s &=&(-1)^s \hf \displaystyle\int {\rm d}^8z\, E^{-1} \Big\{ ~{1\over8}
H'^\rr {\cD}^\b (\cDB^2-4\mu) {\cD}_\b H'_\rr {~~~~~~~~~~~~~~~~}\cr
& &{~~~~~~~~~~~~~~}+\displaystyle \frac{s}{s+1} H'^\rr \big( {\cD}^\a
\cDB^{\ad} G_\ss - \cDB^{\ad} {\cD}^\a \bar G_\ss \big) \cr
& &{~~~~~~~~~~~~~~}+ \displaystyle\frac{(s+1)^2}2
\bar\mu\mu H'\cdot H' +2\bar G\cdot G + \frac{s}{s+1}(G\cdot G+\bar
G\cdot\bar G)~ \Big\} ~~.
\label{13}
\eea
Here ${\rm d}^8z E^{-1}$ is the measure of the AdS superspace.  Any of
the models (\ref{12}) and (\ref{13}) describes the massless on-shell
$osp(1,4)$-representation of the corresponding superspin and its conjugate
representation. The action $S^\bot_{s+1/2}$ is invariant under the
following gauge transformations
\bea
\d H(s,s) &=& g(s,s) + \bar g(s,s) ~~, \non \\
\d \G_\rr &=&\frac s{2(s+1)}
\cDB^{\ad}{\cD}^\a \bar g_\ss ~~.
\label{15}
\eea
with a longitudinal linear parameter $g(s,s)$.   The action $S^\|_{s}$ is
invariant under the following gauge transformations
\bea
\d H'(s-1,s-1) &=& \g(s-1,s-1) + \bar\g(s-1,s-1) ~~, \non \\
\d G_\ss &=& \hf \cDB_\ad \cD_\a \bar \g_\rr ~~.
\label{17}
\eea
with a transversal linear parameter $\g(s,s)$.   It was mentioned in
\cite{ks2} that the linear dynamical variables $\G$, $G$ and their gauge
parameters $g$, $\g$ can always be re-expressed via unconstrained
superfields at the cost of introducing an additional gauge freedom, for
both formulations (\ref{12}) and (\ref{13}), of infinite stage of
reducibility. We are going to show that there exists a generating
formulation in terms of unconstrained superfields for the unified theory
of massless multiplets of all superspins with the action
\be
S = S_0 + \sum_{s=1}^{\infty} S^\|_{s} + S_{\hf} +
\sum_{s=1}^{\infty} S^\bot_{s+\hf} ~~.
\label{18}
\ee
Here the scalar multiplet (superspin-0) is ordinarily described by a
non-gauge (anti)chiral scalar superfield
\be
\cV_0 = \{G, \bar{G}\} ~~, \qquad \qquad \cDB_{\ad}G = 0 ~~,
\label{19}
\ee
with the action
\be
S_0 = \displaystyle\int {\rm d}^8z\, E^{-1} \bar{G}G ~~.
\label{20}
\ee
We also make use of the standard description of the vector multiplet
(superspin-1/2) by a real scalar superfield
\be
\cV_{\hf} = \{H\} ~~,
\label{21}
\ee
with the action
\be
S_{\hf} =\frac{1}{16} \displaystyle\int {\rm d}^8z\, E^{-1}
H {\cD}^\a (\cDB^2-4\m) {\cD}_\a H ~~,
\label{22}
\ee
and the gauge invariance
\be
\d H = g+\bar g, \qquad \qquad \cDB_\ad g = 0 ~~.
\label{23}
\ee
An important observation suggesting the existence of such a formulation
is based on a property of the AdS superspace that a complex tensor
superfield $V(k,l)$ possess a unique decomposition into its transversal and
longitudinal parts
\be
U(k,l)=\G(k,l)+G(k,l) ~~.
\label{8}
\ee
As a consequence, one can equivalently convert the $constrained$ dynamical
variables $\G(s,s)$ (from the superspin-$(s+3/2)$ multiplet) and $G(s,s)$
(from the superspin-$s$ multiplet) into an $unconstrained$ complex superfield
$U(s,s)$ and completely analogous for the gauge parameters $\g(s,s)$ and
$g(s,s)$.

\sect{Generating Formulation}

${~~~~}$ The system with action (\ref{18}) can be reformulated as a gauge
theory in AdS superspace with all the dynamical superfields as well as
gauge parameters taking their values in the algebra of analytic functions
over the one-sheeted hyperboloid in ${\bf R}^{3,1}$
\be
q^a q_a = -\hf q^{\aa} q_{\aa} = 1 ~~.
\label{26}
\ee
An analytic function $\f(q)$ is completely specified by a set of
tensor coefficients $\f_{\ss}$
(symmetric in undotted and dotted indices sepatately), $s=0,1,\ldots$,
that originate in the power series
\be
\f(q) = \sum_{s=0}^{\infty} \f_{\ss} q^{\ss} ~~~,
\label{27}
\ee
where
\be
q^{\ss} =\underbrace{ q^{\aa} \cdots q^{\aa}}_{s \;\; times} ~~.
\label{28}
\ee
The algebra possesses the unique, modulo normalization, Lorentz-invariant
trace
\be
{\rm tr}\;\f = \f(0) \;\; \Longleftrightarrow \;\;
{\rm tr}\;(\f \cdot \j) = \sum_{s=0}^{\infty} \frac{(-1)^s}{s+1}\;
\f^{\ss} \j_{\ss} ~~,
\label{29}
\ee
for arbitrary analytic functions $\f(q)$ and $\j(q)$. Upon introducing the
second-order differential operators
\be
\cQ = q^{\aa} \cD_{\a}\cDB_{\ad} ~~, \qquad
\bar{\cQ} = - q^{\aa} \cDB_{\ad}\cD_{\a} ~~,
\label{30}
\ee
one finds the identity
\be
{\rm tr}\,\displaystyle\int {\rm d}^8z\, E^{-1} \f\cQ \j=
{\rm tr}\,\displaystyle\int {\rm d}^8z\, E^{-1} (\bar{\cQ} \f) \j ~~,
\label{31}
\ee
is valid for arbitrary superfields $\f(z,q)$ and $\j(z,q)$.

Let us consider a linearized theory constructed from an $unconstrained$ complex
superfield $U(z,q)$ and real superfields $X(z,q)$, $Y(z,q)$ which all appear
in the action functional
\bea
S &=& \hf {\rm tr}\;\displaystyle\int {\rm d}^8z\, E^{-1} \Big\{ ~ \frac{1}{4}
Y\left(\bar{\cQ}\cQ + 4(\bar{\cP}-1) \m \bar{\m}\right) X -\frac{\m
\bar{\m}}2 Y (\bar{\cP}-1)^2 Y {~~~~~~~~~~~~~~~~~~} \non \\
& &{~~~~~~~~~~~~~~~~~~~~~}- \frac{\m \bar{\m}}2 X^2 - \hf Y (\cQ \bar{\cP}
U + \bar{\cQ} \cP \bar{U}) + \hf X (\cQ U + \bar{\cQ} \bar{U}) \non \\
& &{~~~~~~~~~~~~~~~~~~~~~}- \bar{U} (\cP + \bar{\cP} - 2) U - U \bar{\cP}
U - \bar{U} \cP \bar{U} ~ \Big\} ~~.
\label{32}
\eea
This action remains unchanged under the following gauge transformations
\bea
\d U &=& - \hf \bar{\cQ}\bar{\ve} ~~, \qquad
\d \bar{U} = - \hf \cQ \ve ~~, \non \\
\d Y &=& \ve + \bar{\ve} ~~, \qquad
\d X = (\bar{\cP} - 1) \ve +(\cP - 1) \bar{\ve} ~~,
\label{33}
\eea
expressed in terms of an $unconstrained$ complex gauge parameter $\ve(z,q)$.
The corresponding generators are obviously seen to be linearly independent.

The actions (\ref{18}) and (\ref{32}) can be used to define one and the same
physical theory!  This can be proven by making use of the following
decompositions
\bea
U &=& \sum_{s=0}^{\infty} U_{\ss} q^{\ss} ~~, \quad
U(s,s) = \G(s,s) + G(s,s) ~~;
\label{34} \\
Y &=& \sum_{s=0}^{\infty} Y_{\ss} q^{\ss} ~~, \quad
Y(s,s) = H'(s,s) - H(s,s) ~~;
\label{35} \\
X &=& \sum_{s=0}^{\infty} X_{\ss} q^{\ss} ~~, \quad
X(s,s) = (s+1) \{H'(s,s) + H(s,s)\} ~~;
\label{36}
\eea
where (\ref{8}) has been applied to (3.9). Now, calculating the trace
in (\ref{32}) reduces it to that given by (\ref{18}). Similarly,
representing the gauge parameter in the form
\be
\ve = \sum_{s=0}^{\infty} \ve_{\ss} q^{\ss} ~~, \quad
\ve(s,s) = \g(s,s) - g(s,s) ~~,
\label{37}
\ee
equation (\ref{33}) proves to be equivalent to the gauge transformations
(\ref{15}), (\ref{17}) and (\ref{23}).

As can be seen from (\ref{35}) and (\ref{36}), the expansions of the
real $q$-analytic superfields $X$ and $Y$ are expressed in terms of
the coefficients $H'(s,s)$ and $H(s,s)$.  This in turns implies that
we may define two additional $q$-analytic superfields via the equations
\bea
{\cal A}' &=& \sum_{s=0}^{\infty} {H'}_{\ss} q^{\ss} ~~,
\label{201} \\
{\cal A} &=& \sum_{s=0}^{\infty} {H}_{\ss} q^{\ss} ~~.
\label{202}
\eea
Consequently, we find the results
\be
Y ~\equiv ~ {\cal A}' ~-~ {\cal A}  ~~~,~~~ X ~\equiv~ \Big[ ~ 1 \, +
\, q^{\a \ad} ~ {{\partial {~~~} \over\partial q^{\a \ad} }} ~  \Big]
[ \, {\cal A}' ~+~ {\cal A} \, ] ~~.
\label{203}
\ee

A remarkable feature of the theory in field is that it admits the
simple gauge condition
\be
Y = 0 \qquad \Rightarrow \qquad \ve = i\rho ~~, \quad \rho = \bar{\rho} ~~.
\label{38}
\ee
and in this gauge the action reduces to the form
\bea
S &=& \hf {\rm tr}\;\displaystyle\int {\rm d}^8z\, E^{-1}  \Big\{
-\frac{\m \bar{\m}}2 X^2 + \hf X (\cQ U + \bar{\cQ} \bar{U}){~~~~~~~~~~
~~~~~~~}\non \\
& &{~~~~~~~~~~~~~~~~~~~~~}- \bar{U} (\cP + \bar{\cP} - 2) U - U
\bar{\cP} U - \bar{U} \cP \bar{U} ~ \Big\}
\label{39}
\eea
and the residual gauge transformations takes the forms
\bea
\d X &=& i (\bar{\cP} - \cP) \rho = - \frac{i}{2\m \bar{\m}}
[\cQ, \bar{\cQ}] \rho ~~, \non \\
\d U &=&  \frac i2  \bar{\cQ}\rho , \qquad
\d \bar{U} = - \frac i2 \cQ \rho ~~,
\label{40}
\eea
with $\rho$ being a real unconstrained superfield. Here we have used the
identity
\be
[\cQ,\bar{\cQ}] = 2\m \bar{\m} (\cP - \bar{\cP}).
\label{41}
\ee

An additional surprise is that $X$ is nothing more but an auxiliary
superfield. In the gauge (\ref{38}) this can be eliminated with the
aid of the equation of motion
\be
X = - \frac{1}{\m \bar{\m}} (\cQ U + \bar{\cQ} \bar{U}) ~~.
\label{42}
\ee
As a result, the theory is completely described by the complex
unconstrained superfield $U(z,q)$.

   It worth pointing out that any complex superfield has a unique
decomposition into the sum of transversal and longitudinal linear ones
(see (\ref{34}) and (\ref{37})) only in the AdS superspace at $\mu\neq 0$.
Nevertheless the action (\ref{39}) has a well defined flat limit. To see
this, let us set $\mu=\bar\mu$ and make the replacements
$$
U\Rightarrow \sqrt{\mu} U, \qquad X\Rightarrow X/\sqrt{\mu}, \qquad
\rho\Rightarrow \sqrt{\mu} \rho
$$
in (\ref{39}) and (\ref{40}). Due to these re-definitions all the singularities
at $\mu=0$ disappear and in this limit we arrive at the action
\be
S = \frac14 {\rm tr}\;\displaystyle\int {\rm d}^8z\,  \Big\{
X (Q U + \bar Q \bar{U})
- \bar{U} (D^2 + \bar D^2) U - U \bar D^2 U - \bar{U} D^2 \bar{U}
\Big\}~~,
\ee
and the gauge transformations
\be
\d X = \frac i2 (\bar D^2 - D^2) \rho~~~, \qquad
\d U =  \frac i2  \bar{Q}\rho , \qquad
\d \bar{U} = - \frac i2 Q \rho~~,
\label{40}
\ee
(here $Q = q^\aa D_\a \bar D_\ad$) in flat superspace.  Note that the
superfield $X$ is no longer auxiliary.  Although (3.21) is the formal
Minkowski space limit of our anti-de-Sitter theory, it will require
additional investigation before its full relevance is understood.

It can be seen from (\ref{35}) that the condition $Y = 0$ is equivalent to
the condition $H = H'$.  We may ask what is the significance of this?  From
(\ref{10}) and (\ref{11}), it should be clear that in the gauge $Y = 0$, the
physical content of both the superspin-s and superspin-(s + 1/2) multiplets
are realized as component fields in a single real superfield $H$. In the
superspace supergravity theory \cite{gmov} associated with the $N = 1$, $D
= 4$ heterotic string a similar phenomenon is known to occur.  In the
special $\b$FFC supergeometry of \cite{gmov}, the graviton supermultiplet
as well as axion supermultiplet {\it {all}} occur as components of the
usual axial-vector supergravity prepotential $H_{\a \ad}$.  The special
$\b$FFC supergeometry is also known in the literature as ``the string gauge.''
Thus, we conjecture that here the condition $Y = 0$ may be the analog
of the string gauge. It should also be clear that the q-analytic
superfield ${\cal A} (z,q)$ plays a role analogous to the string-field
functional in linearized covariant string field theory.

\sect{Global Symmetries}

${~~~~}$It follows from (\ref{40}) that the parameters of global symmetries
of the unified model should solve of the equation
\be
\cQ \U = 0 ~~, \quad \U = \bar{\U} ~~, \qquad \Rightarrow \qquad
(\cP - \bar\cP) \U = q^{\aa} \cD_{\aa} \U = 0 ~~.
\label{43}
\ee
This proves to have the following general solution
\be
\U = \sum_{s=0}^{\infty} \{ t_{\ss} + l_{\ss} \} q^{\ss} ~~,
\label{44}
\ee
where the transversal $t$'s and longitudinal $l$'s coefficients
are constrained by
\be
t_{\ss} = \bar{t}_{\ss} ~~, \qquad \cDB^\ad t_{\ss} =
\cD_\a \cDB _\ad t_{\ss} = 0 ~~,
\label{45}
\ee
and
\be
l_{\ss} = \bar{l}_{\ss} ~~, \qquad \cDB_\ad l_{\ss} =
\cD^\a \cDB ^\ad l_{\ss} = 0 ~~,
\label{46}
\ee
respectively. The constraints in (\ref{45}) and (\ref{46}) coincide
with Killing equations for tensor superfields and define a complete set
of independent Killing tensor superfields. In the AdS superspace their
solutions are parametrized by a finite number of constant parameters.
The component content of the Killing superfields $t(s,s)$ and $l(s,s)$
is given by the Killing tensor fields
\be
\begin{array}{ll}
\ve_\ss = t_\ss|~~~, & \qquad \ve_\ts = \cD_\a t_\ss|~~~, \\
\ve_\ss = l_\ss|~~~, & \qquad \ve_\rs = \cD^\a l_\ss|~~~,
\end{array}
\ee
satisfying the conditions
$$
\nabla^\aa \ve_\kl = 0 ~~, \quad \nabla_\aa \ve_\kl = 0~~~.
$$
Here the symbol `$|$' indicates the zeroth-order component in $\q=0$,
$\nabla$'s denote the covariant derivatives of the AdS space.

The space of Killing tensor superfields (\ref{45}), (\ref{46}) may be
endowed with the structure of a superalgebra. Given two tensors,
for example both transversal $t(s,s)$ and $t(r,r)$, one can
build a Killing longitudinal tensor $l(s+r-2p+1,s+r-2p+1)$ for $p =
0,1,\ldots,\min{(s,r)}$ as their antisymmetric bilinear combination
defined uniquely modulo a factor from the equations (\ref{45}).
In general the following combinations are available
\be
\begin{array}{lllll}
l(s,s) & {\rm and} & l(r,r) & {\rm combine~to} & l(s+r-2p-1,s+r-2p-1)~~~,\\
&&&&p=0,1,\ldots,\min{(s-1,r-1)}~~~, \\
t(s,s) & {\rm and} & t(r,r) & {\rm combine~to} & l(s+r-2p+1,s+r-2p+1)~~~,\\
&&&&p=0,1,\ldots,\min{(s,r)}~~~, \\
l(s,s) & {\rm and} & t(r,r) & {\rm combine~to} & t(s+r-2p-1,s+r-2p-1)~~~,\\
&&&&p=0,1,\ldots,\min{(s-1,r)}~~~.
\end{array}
\label{47}\ee
Here we present manifestly only the first sector. For that purpose
let us introduce the notations $l^n(s+n,s-n)$,
$n=-s,-s+1,\ldots,s$, and $l^{n+1/2}(s+n,s-n-1)$,
$n=-s,-s+1,\ldots,s-1$, $(l^\nu)^* = l^{-\nu}$
for the derivatives of the longitudinal Killing
tensor $l^0(s,s) \equiv l(s,s)$ by the recurrent relations
\bea
%&l^0_\ss = l_\ss, \qquad (l^\nu)^* = l^{-\nu}~~~, \cr
&{\cD_\a}^\ad l^n_{\a(s+n)\ad(s-n)} = i(s-n+1) \mu
l^{n+1}_{\a(s+n+1)\ad(s-n-1)}~~~, \\
&\cDB^\ad l^n_{\a(s+n)\ad(s-n)} = (s-n+1) \mu
l^{n+1/2}_{\a(s+n)\ad(s-n-1)} ~~~, \\
&\cDB_\ad l^{n+1/2}_{\a(s+n)\ad(s-n-1)} =
-2 l^n_{\a(s+n)\ad(s-n)} ~~~,
\eea
and their complex conjugates. Now if $l(s,s)$ and $l'(r,r)$ are two
longitudinal Killing superfields, any longitudinal Killing superfield
with $s+r-2p-1$ undotted and dotted indices ($p=0,1,\ldots,\min{(s-1,r-1)}$)
that can be built of $l$ and $l'$ is proportional to
\bea
&& l''_{\a(s+r-2p-1)\ad(s+r-2p-1)} = \cr
&&{~~~}\sum\limits_q \b_{q+1/2} \Big(\sum\limits_{n=-p}^p \a_n^{q+1/2}
l_{\a(s-p+q)\ad(s-p-q-1)}^{n+q+1/2\:\b(p+n)\bd(p-n)}
 l'^{n-q-1/2}_{\a(r-p-q-1)\b(p+n)\ad(r-p+q)\bd(p-n)} \cr
&&{~~~}+ \sum\limits_{n=-p-1}^p \a_{n+1/2}^{q+1/2}
l_{\a(s-p+q)\ad(s-p-q-1)}^{n+q+1\:\b(p+n+1)\bd(p-n)}
l'^{n-q}_{\a(r-p-q-1)\b(p+n+1)\ad(r-p+q)\bd(p-n)} \Big)
\label{51}\eea
where the summation over $q$ under the conditions $s-p-1 \ge q \ge p-s$
and $r-p-1 \ge q \ge p-r$ and the constants $\b$ and $\a$ are equal
\bea
\a_n^{q+1/2} &=& \frac{i(-1)^n}{(p+n)!(p-n)!} \frac{\mu^n}{\bar\mu^n}
~~, \qquad \a_{n+1/2}^{q+1/2} = \frac{2i(-1)^{n+1}}{(p+n+1)!(p-n)!}
\frac{\mu^n}{\bar\mu^{n+1}}~~~, \cr
\b_{q+1/2} &=& \big[(s-p+q)!(s-p-q-1)!(r-p+q)!(r-p-q-1)!\big]^{-1}~~~.
\non\eea
They satisfy the relations
$$
\bar\b_{q+1/2} = \b_{-q-1/2} ~~, \qquad \bar\a_n^{q+1/2} = - \a_{-n}^{-q-1/2}
~~, \qquad \bar\a_{n+1/2}^{q+1/2} = \a_{-n-1/2}^{-q-1/2} ~~,
$$
that ensure simultaneously the reality of $l''$ and its antisymmetricity
with respect to the replacement $l(s,s)\leftrightarrow l'(r,r)$.

   Let us consider the lowest components of the Killing superfield $\U$,
namely $t$ and $l_\aa$.  They prove to describe important symmetries of the
model: $N=1$ AdS transformations (see, e.g., \cite{bk,ggrs}) and $N=2$ global
supersymmetry \cite{gks} respectively.

    The $N=1$ AdS transformations can be written in the form
\be
\d X = \cK X ~~, \qquad \d Y = \cK Y ~~, \qquad \d U = \cK U~~~,
\label{52}\ee
where the operator $\cK$ is expressed in terms of special Killing superfield
$\U$:
\be
\cK = \hf q^{\aa} \left\{ \frac i2 (\cD_\a \U) \cDB_{\ad} + {\rm c.c.}
+ ({M_\a}^\b \U) \cD_{\b\ad} - (\cD_{\b\ad} \U) {M^\b}_\a
+ \hf \U \cD_\aa \right\}, \; .
\ee
where $\U \equiv q^{\aa} k_{\aa}$.  Another symmetry of the action (\ref{39})
involves the scalar transversal component $t$ of the Killing superfield $\U$:
\bea
\d_t Y &=& 0~~~,\qquad
\d_t X = 2i [\bar{\cP}, t]\, \cR U + {\rm c.c.}~~~, \cr
\d_t U &=& \frac i2 \m \bar{\m} [\cP + \bar{\cP}, t]\, \cR X
+ \frac i2 \m\bar{\m} (\bar\cP-1) [\cP + \bar{\cP}, t]\, \cR Y \cr
&&+ i q^{\aa} (\cD_\a t) \cDB_{\ad} \cR (U + \bar{U})~~~, \label{53}
\eea
where $R$ denotes the reflection on the hyperboloid
\be
\cR \f (q) = \f (-q) ~~.
\label{54}\ee
These are exactly the $N=2$ supersymmetry transformations and $O(2)$ rotations
of the $N=2$ AdS superalgebra, which were found in components in \cite{gks}.
Therefore we conclude that the action (\ref{39}) possesses $N=2$ AdS
supersymmetry which is described in the covariant form in terms of the
generating superfields $X(z,q)$, $Y(z,q)$ and $U(z,q)$ with the Killing
parameter $\U(z,q)$ containing only $t$ and $l_\aa$ components.
The $N=2$ transformations turn out to form a closed
algebra off the mass-shell.  In the gauge $Y=0$ a commutator of two
transformations (\ref{53}) is given by the sum  of an AdS transformation
(\ref{52})  and a gauge variation:
\bea
\left[ \delta_t , \delta_{t'}\right] X &=& \cK X
+ i(\bar{\cP} - \cP) \rho ~~, \non \\
\left[ \delta_t , \delta_{t'}\right] U &=& \cK U
+ \frac{i}{2} \bar{\cQ} \rho ~~,
\label{55}\eea
where $\cK$ is given by (\ref{52}) with
\be
\U = 4 i \left((\cD_\a t) (\bar\cD_\ad t') - (\cD_\a t) (\bar\cD_\ad
t')\right)q^\aa ~~, \qquad \rho = \hf \U (U + \bar{U}) ~~.
\label{56}\ee

    In the work of \cite{fv} the superalgebra of higher spins and auxiliary
fields $shsa(1)$ was proposed to play the role of the structure algebra of a
higher spin theory. This algebra describes the global symmetries of the
free action as well. As the spin content of our model coincides with
that arising in the framework of \cite{vas1}, the superalgebra $shsa(1)$
can be considered as a global symmetry algebra of the action (\ref{39})
too.  It is then natural to identify the $N=2$ AdS superalgebra of the
transformations (\ref{52}) and (\ref{53}) with the finite-dimensional
subalgebra of $shsa(1)$.   The fact that these transformations are expressed
in terms of Killing superfields along with some observations at the component
level allows us to suggest that all the global symmetries can be expressed
in terms of Killing tensors (\ref{45}) and (\ref{46}). Then the
(anti)commutator relations of the superalgebra of global symmetries
could be described by the bilinear combinations (\ref{47}), (\ref{51})
after appropriate normalization of the multiplicative factors. Most likely
the Killing tensors entering the superfield $\U$ (\ref{51}) correspond
only to the physical subalgebra of $shsa(1)$ (at equal powers of the Klein
operators \cite{fv}) because the number and the Lorentz structure of
generators in this subalgebra coincides with those of the constant
parameters in the solutions of (\ref{45}) and (\ref{46}).

In closing this work, we also note that the issue of duality can also
be investigated within the context of the universal action in (\ref{32})
as there seems to be no fundamental impediment to performing a
superfield duality transformation upon this action.  Presumably such
a duality transformation would replace the chiral scalar multiplet
described in (\ref{19}) and (\ref{20}) by a chiral spinor superfield
since the latter is known to describe the axion multiplet in the
work on the $N = 1$, $D = 4$ supergeometry \cite{gmov} that arises
from the heterotic string. 
It would be an interesting exercise to show that the duality
transformation when implemented on the generating formalism, does
indeed interchange the two dually equivalent superfield 
formulations \cite{ks2}. But then it is worth expecting the appearance,
in the generating formalism, of inverse powers of $(2\cP - 1)$ which
turn into numerical factors only upon 
passing to the components in $q$ and decomposing the 
complex unconstrained superfields into their transversal and longitudinal
parts by the rule (\ref{8}). 
\vspace{1cm}

\noindent
{\bf Acknowledgements}\\
One of us (SMK) thanks the Institute for Theoretical Physics at
Hannover, where the present paper was completed, for hospitality.
This work was supported in part by the Russain Foundation for Basic
Research Grant No. 96-02-16017 and the Joint DFG--RFBR Project
Grant No. 96-02-00180G. The research of SJG was supported by the
U.S. NSF under grant PHY-96-43219 and NATO under grant CRG-930789.
The work of AGS was also supported by a Fellowship of Tomalla
Foundation (under the program of the Interanational Center for
Fundamental Physics in Moscow).

\end{document}